 \documentclass[12pt]{article}                  
\usepackage{graphicx}
\usepackage{psfrag}
\usepackage{amsmath,  amssymb}
\usepackage{mathptmx}

\newcommand{\A}{{\cal A}}
\newcommand{\M}{{\cal M}}
\newcommand{\PP}{{\cal P}}
\newcommand{\eps}{{\varepsilon}}
\newcommand{\ch}{{\mbox{\rm ch}}}
\newcommand{\smsp}{\hspace{0.3mm}}
\newcommand{\e}{\mathbb{E}}

\newcommand{\Reals}{\mathbb{R}}
\newcommand{\Natural}{\mathbb{N}}
\newcommand{\la}{\langle}
\newcommand{\ra}{\rangle}

\begin{document}

\title{The Sherrington-Kirkpatrick model: an overview}
\author{Dmitry Panchenko\thanks{Department of Mathematics, Texas A\&M University, email: panchenk@math.tamu.edu. 
Partially supported by NSF grant.}\\
}
\maketitle
\begin{abstract}
The goal of this paper is to review some of the main ideas that emerged from the attempts to confirm mathematically the predictions of the celebrated Parisi ansatz in the Sherrington-Kirkpatrick model. We try to focus on the big picture while sketching the proofs of only a few selected results, but an interested reader can find most of the missing details in \cite{SKmodel} and \cite{SG2}.
\end{abstract}
\vspace{0.5cm}
Key words: Sherrington-Kirkpatrick model, Parisi ansatz.\\
Mathematics Subject Classification (2010): 60K35,  82B44

\section{Introduction}

\paragraph{The Sherrington-Kirkpatrick model.} In $1975$, Sherrington and Kirkpatrick \cite{SK} introduced a mean field model for a spin glass---a disordered magnetic alloy that exhibits unusual magnetic behavior. Given a configuration of $N$ Ising {spins},
$$
\sigma = (\sigma_1,\ldots,\sigma_N) \in \varSigma_N = \{-1,+1\}^N,
$$
the {Hamiltonian} of the model is given by
\begin{equation}
H_N(\sigma) = \frac{1}{\sqrt{N}} \sum_{i,j =1}^N g_{ij}\sigma_i \sigma_j,
\label{SKH}
\end{equation}
where $(g_{ij})$ are i.i.d. standard Gaussian random variables, collectively called the {disorder} of the model. The fact that the distribution of $H_N(\sigma)$ is invariant under the permutations of the coordinates of $\sigma$ is called the symmetry between sites, which is what one usually understands by a mean field model. The Hamiltonian (\ref{SKH}) is a Gaussian process with the covariance 
\begin{equation}
\e H_N(\sigma^1)H_N(\sigma^2)
=
\frac{1}{N} \sum_{i,j =1}^N \sigma_i^1 \sigma_j^1  \sigma_i^2 \sigma_j^2
=
N\Bigl(
\frac{1}{N} \sum_{i=1}^N \sigma_i^1  \sigma_i^2 
\Bigr)^2
=
N R_{1,2}^2
\label{Cov}
\end{equation}
that depends on the spin configurations $\sigma^1, \sigma^2$ only through their normalized scalar product
\begin{equation}
R_{1,2} = \frac{1}{N} \sigma^1 \cdot \sigma^2 
=\frac{1}{N} \sum_{i=1}^N \sigma_i^1  \sigma_i^2, 
\label{overlap}
\end{equation}
called the {overlap} of $\sigma^1$ and $\sigma^2$. Since the distribution of a Gaussian process is determined by its covariance, it is not surprising that the overlaps play a central role in the analysis of the model. One can also consider a generalization of the Sherrington-Kirkpatrick model, the so-called {mixed $p$-spin model}, which corresponds to the Hamiltonian 
\begin{equation}
H_N(\sigma) = \sum_{p\geq 1} \beta_p H_{N,p}(\sigma)
\label{mixedH}
\end{equation}
given by a linear combination of {pure $p$-spin} Hamiltonians\index{Hamiltonian!$p$-spin model}
\begin{equation}
H_{N,p}(\sigma)
=
\frac{1}{N^{(p-1)/2}}
\sum_{i_1,\ldots,i_p = 1}^N g_{i_1\ldots i_p} \sigma_{i_1}\cdots\sigma_{i_p},
\label{mixedp}
\end{equation}
where the random variables $(g_{i_1\ldots i_p})$ are standard Gaussian, independent for all $p\geq 1$ and all $(i_1,\ldots,i_p)$. Similarly to (\ref{Cov}), it is easy to check that the covariance is, again, a function of the overlap,
\begin{equation}
\e H_N(\sigma^1) H_N(\sigma^2) = N\xi(R_{1,2}),
\,\mbox{ where }\, \xi(x)=\sum_{p\geq 1}\beta_p^2 x^p.
\label{Covxi}
\end{equation}
One usually assumes that the coefficients $(\beta_p)$ decrease fast enough to ensure that the process is well defined when the sum in (\ref{mixedH}) includes infinitely many terms. The model may also include the {external field} term $h(\sigma_1+\ldots+\sigma_N)$ with the external field parameter $h\in\Reals.$ For simplicity of notation, we will assume that $h=0$, but all the results hold in the presence of the external field with some minor modifications. 
One of the main problems in these models is to understand the behavior of the {ground state energy} $\min_{\sigma\in\Sigma_N} H_N(\sigma)$ in the {thermodynamic limit} $N\to\infty$. In a standard way this problem can be reduced to the computation of the limit of the {free energy}
\begin{equation}
F_N = \frac{1}{N}\smsp \e \log Z_N,
\,\mbox{ where }\,
Z_N = \sum_{\sigma\in\Sigma_N } \exp \bigl(- \beta H_N(\sigma) \bigr),
\label{FEZN}
\end{equation}
for each {inverse temperature parameter} $\beta = 1/T>0$, and a formula for this limit was proposed by Sherrington and Kirkpatrick in \cite{SK} based on the so-called replica formalism. At the same time, they observed that their {replica symmetric solution} exhibits ``unphysical behavior" at low temperature, which means that it can only be correct at high temperature. Several years later, Parisi proposed in \cite{Parisi79}, \cite{Parisi}, another, {replica symmetry breaking, solution} within replica theory, now called the {Parisi ansatz}, which was consistent at any temperature $T> 0$ and, moreover, was in excellent agreement with computer simulations. The Parisi formula for the free energy is given by the following variational principle. 

\paragraph{The Parisi formula.} A basic parameter, called the {functional order parameter}, is a distribution function $\zeta$ on $[0,1]$,
\begin{equation}
\zeta\bigl(\bigl\{q_p\bigr\}\bigr) = \zeta_{p} - \zeta_{p-1}
\,\mbox{ for }\,
p=0,\ldots, r,
\label{ch30zetafop}
\end{equation}
corresponding to the choice of $r\geq 1$,
\begin{equation}
0=\zeta_{-1}< \zeta_0 <\ldots < \zeta_{r-1} < \zeta_r = 1
\label{ch31zetas}
\end{equation}
and
\begin{equation}
0=q_0<q_1 <\ldots <q_{r-1}< q_r =1.
\label{ch31qs}
\end{equation}
Notice that $\zeta$ carries some weight on $r-1$ points inside the interval $(0,1)$ and on the points $0$ and $1$. In general, one can remove the atoms $q_0 = 0$ and $q_r = 1$ and allow $\zeta_0 = 0$ and $\zeta_{r-1}=1$, but these cases can be recovered by continuity, so it will be convenient to assume that the inequalities in (\ref{ch31zetas}) are strict. Next, we consider i.i.d. standard Gaussian random variables  $(\eta_p)_{0\leq p\leq r}$ 
and define 
\begin{equation}
X_r
=
 \log  \ch\smsp \beta\Bigl(
 \eta_0 \xi'(0)^{1/2} +
\sum_{1\leq p\leq r} \eta_p \bigl(\xi'(q_{p}) - \xi'(q_{p-1}) \bigr)^{1/2}
\Bigr).
\label{ch30Xr}
\end{equation}
Recursively over $0\leq l\leq r-1,$ we define
\begin{equation}
X_l=\frac{1}{\zeta_l}\log \e_l\exp \zeta_l X_{l+1},
\label{ch30Xl}
\end{equation}
where $\e_l$ denotes the expectation with respect to $\eta_{l+1}$ only. Notice that $X_0$ is a function of $\eta_0$. Finally, we let $\theta(x) = x \xi'(x) - \xi(x)$ and define the so-called  Parisi functional by
\begin{equation}
\PP(\zeta)
=
\log 2 +
\e X_0
-\frac{1}{2}
\sum_{0\leq p\leq r-1} \zeta_p \bigl(\theta(q_{p+1}) - \theta(q_{p})\bigr).
\label{ch30Pzeta}
\end{equation}
The Parisi solution predicted that the limit of the free energy is equal to
\begin{equation}
\lim_{N\to\infty} F_N
= 
\inf_\zeta \PP(\zeta),
\label{ch30Parisi}
\end{equation}
where the infimum is taken over all distribution functions $\zeta$ as above or, in other words, over all $r\geq 1$ and sequences (\ref{ch31zetas}) and (\ref{ch31qs}). The replica method by which the formula (\ref{ch30Parisi}) was discovered did not give a definite interpretation of the f.o.p. $\zeta$ or the functional (\ref{ch30Pzeta}), but a more clear picture emerged in the physics literature (a classical reference is \cite{MPV}) during the subsequent interpretation of the Parisi ansatz in terms of some physical properties of the {Gibbs measure} of the model,
\begin{equation}
G_N(\sigma) = \frac{\exp(- \beta H_N(\sigma))}{Z_N},
\end{equation}
where the normalizing factor $Z_N$ defined in (\ref{FEZN}) is called the partition function. To describe this picture, let us first explain a modern mathematical framework that is used to encode relevant information about the model in the thermodynamic limit. 

\paragraph{Asymptotic Gibbs' measures.} Notice that, due to the special covariance structure (\ref{Covxi}), the distribution of the Gaussian Hamiltonian (\ref{mixedH}) is invariant under orthogonal transformations of the set of spin configurations $\Sigma_N$, which means that, given any orthogonal transformation $U$ on $\Reals^N$, we have the equality in distribution
$$
\bigl( H_N(U(\sigma))\bigr)_{\sigma\in \Sigma_N}
\stackrel{d}{=}
\bigl( H_N(\sigma)\bigr)_{\sigma\in \Sigma_N}.
$$
As a result, we are just as interested in the measure $G_N\circ U^{-1}$ on the set $U(\Sigma_N)$ as in the original Gibbs measure $G_N$. To encode the information about $G_N$ up to orthogonal transformations, let us consider an i.i.d. sequence $(\sigma^l)_{l\geq 1}$ of replicas sampled from $G_N$ and consider the normalized Gram matrix of their overlaps 
\begin{equation}
R^N= \bigl(R^N_{l,l'}\bigr)_{l,l'\geq 1} = \frac{1}{N}\bigl(\sigma^l \cdot \sigma^{l'}\bigr)_{l,l'\geq 1}.
\label{ch12overlaps}
\end{equation}
It is easy to see that, given $R^N$, one can reconstruct the Gibbs measure $G_N$ up to orthogonal transformations, because, every time we observe an entry $R_{l,l'}$ equal to $1$, it means that the replicas $l$ and $l'$ are equal. This way we can group equal replicas and then use the law of large numbers to estimate their Gibbs weights from the frequencies of their appearance in the sample. Since the Gram matrix describes relative position of points in the Euclidean space up to orthogonal transformations, the overlap matrix $R^N$ can be used to encode the information about the Gibbs measure $G_N$ up to orthogonal transformations. For this reason (and some other reasons that will be mentioned below), the Gibbs measure in the Sherrington-Kirkpatrick and mixed $p$-spin models is often identified with the distribution of the overlap matrix $R^N$. Since the overlaps are bounded in absolute value by $1$, this allows us to pass to the infinite-volume limit and consider a set of all possible limiting distributions of $R^N$ over subsequences. An infinite array $R$ with any such limiting distribution inherits two basic properties of $R^N$. First, it is non-negative definite and, second, it satisfies a ``replica symmetry" property
\begin{equation}
\bigl(R_{\pi(l), \pi(l')}\bigr)_{l,l'\geq 1}  \stackrel{d}{=}\bigl(R_{l,l'}\bigr)_{l,l'\geq 1},
\label{ch12weakexch}
\end{equation}
for any permutation $\pi$ of finitely many indices, where the equality is in distribution. Such arrays are called Gram-de Finetti arrays and the Dovbysh-Sudakov representation \cite{DS} (see also \cite{PDS}) guarantees the existence of a random measure $G$ on the unit ball of a separable Hilbert space $H$ such that
\begin{equation}
\bigl(R_{l,l'}\bigr)_{l\not = l'}\stackrel{d}{=}
\bigl(\sigma^{l}\cdot \sigma^{l'}\bigr)_{l\not = l'},
\label{ch12RDS}
\end{equation}
where $(\sigma^l)$ is an i.i.d. sequence of replicas sampled from the measure $G$. We will call such measure $G$ an {asymptotic Gibbs measure} and think of it as a limit of the Gibbs measures $G_N$ over some subsequence, where the convergence is defined by way of the overlap arrays, as above. The reason why the diagonal elements are not included in (\ref{ch12RDS}) is because in (\ref{ch12overlaps}) they were equal to $1$ by construction, while the asymptotic Gibbs measure is not necessarily concentrated on the unit sphere. This mathematical definition of an asymptotic Gibbs measure via the Dovbysh-Sudakov representation was first given by Arguin and Aizenman in \cite{AA}. We will now describe (reinterpret) various predictions of the Parisi ansatz in the language of these asymptotic Gibbs measures.  

\paragraph{Order parameter and pure states.} First of all, in the work of Parisi, \cite{Parisi83}, the {functional order parameter} $\zeta$ in (\ref{ch30zetafop}) was identified with the distribution of the overlap $R_{1,2}$ under the average (asymptotic) Gibbs measure, 
\begin{equation}
\zeta(A) = \e G^{\otimes 2}\bigl((\sigma^1,\sigma^2) : R_{1,2} = \sigma^1\cdot \sigma^2 \in A\bigr),
\label{Gfop}
\end{equation}
and the infimum in the Parisi formula (\ref{ch30Parisi}) is taken over all possible candidates for this distribution in the thermodynamic limit. The fact that the infimum in (\ref{ch30Parisi}) is taken over discrete distributions $\zeta$ is not critical, since the definition of the Parisi functional $\PP(\zeta)$ can be extended to all distributions on $[0,1]$ by continuity. Another important idea introduced in  \cite{Parisi83} was the decomposition of the Gibbs measure into pure states. This simply means that an asymptotic Gibbs measure $G$ is concentrated on countably many points $(h_l)_{l\geq 1}$ in the Hilbert space $H$, and these are precisely the pure states. It was also suggested in \cite{Parisi83} that it is reasonable to assume that all the pure states have equal norm, so the asymptotic Gibbs measure $G$ is concentrated on some non-random sphere, $G(\|h\| = c) =1$. This implies, for example, that the largest value the overlap can take is $R_{1,2}=c^2$ when the replicas $\sigma^1=\sigma^2 = h_l$ for some $l\geq 1$ and, since this can happen with positive probability, the distribution of the overlap has an atom at the largest point of its support, $\zeta(\{c^2\})>0$. We will see below that, due to some stability properties of the Gibbs measure, the pure state picture is correct if the distribution of the overlap has an atom at the largest point $c^2$ of its support, otherwise, the measure $G$ is non-atomic, but is still concentrated on the sphere $\|h\| = c$. 

\paragraph{Ultrametricity.} Perhaps, the most famous feature of the Parisi solution of the SK model in \cite{Parisi79}, \cite{Parisi}, was the choice of an ultrametric parametrization of the replica matrix in the replica method, and in the work of M\'ezard, Parisi, Sourlas, Toulouse and Virasoro \cite{M1}, \cite{M2}, this was interpreted as the ultrametricity of the support of the asymptotic Gibbs measure $G$ in $H$, which means that the distances between any three points in the support satisfy the strong triangle, or ultrametric, inequality 
\begin{equation}
\|{\sigma}^2 - {\sigma}^3\|
\leq
\max\bigl(\|{\sigma}^1 - {\sigma}^2\|,
\|{\sigma}^1 -{\sigma}^3\|\bigr).
\label{ultra1}
\end{equation}
When the Gibbs measure is concentrated on the sphere $\|h\| = c$, we can express the distance in terms of the overlap, $\|{\sigma}^1 - {\sigma}^2\|^2 = 2(c^2 - R_{1,2})$, and, therefore, the ultrametricity can also be expressed in terms of the overlaps,
\begin{equation}
R_{2,3} \geq \min(R_{1,2}, R_{1,3}).
\label{ultra2}
\end{equation}
One can think about ultrametricity as clustering of the support of $G$, because the ultrametric inequality (\ref{ultra1}) implies that the relation defined by the condition
\begin{equation}
\sigma^1 \sim_d \sigma^2 \Longleftrightarrow \|{\sigma}^1 - {\sigma}^2\| \leq d
\label{diameter}
\end{equation}
is an equivalence relation on the support of $G$ for any $d\geq 0$. As we increase $d$, smaller clusters will collapse into bigger clusters and the whole process can be visualized by a branching tree. For a given diameter $d\geq 0,$ one can consider the equivalence clusters and study the joint distribution of their Gibbs weights. In the case when $d=0$, which corresponds to the weights of the pure states $(G(\{h_l \}))_{l\geq 1}$, this distribution was characterized in \cite{M1} using the replica method, but the same computation works for any $d\geq 0$. More generally, one can consider several cluster sizes $d_1<\ldots < d_r$ and for each pure state $h_l$ consider the weights of the clusters it belongs to,
$$
G(\|\sigma - h_l\| \leq d_1),\ldots,G(\|\sigma - h_l\| \leq d_r). 
$$
Again, using the replica method within the Parisi ansatz, one can study the joint distribution of all these weights for all pure states, but the computation gets very complicated and not particularly illuminating. On the other hand, the problem of understanding the distribution of the cluster weights is very important, since this gives, in some sense, a complete description of the asymptotic Gibbs measure $G$. Fortunately, a much more explicit and useful description of the asymptotic Gibbs measures arose from the study of some related toy models.

\paragraph{Derrida's random energy models.} In the early eighties, Derrida proposed two simplified models of spin glasses: the random energy model (REM) in \cite{DerridaREM1}, \cite{DerridaREM2}, and the generalized random energy model (GREM) in \cite{DerridaGREM}, \cite{DerridaGREM2}. The Hamiltonian of the REM is given by a vector $(H_N(\sigma))_{\sigma\in\Sigma_N}$ of independent Gaussian random variables with variance $N$, which is a rather classical object. The GREM combines several random energy models in a hierarchical way with the ultrametric structure built into the model from the beginning. Even though these simplified models do not shed light on the Parisi ansatz in the SK model directly, the behavior of the Gibbs measures in these models was predicted to be, in some sense, identical to that of the SK model. For example, Derrida and Toulouse showed in \cite{DerridaToulouse} that the Gibbs weights in the REM have the same distribution in the thermodynamic limit as the Gibbs weights of the pure states in the SK model, described in \cite{M1}, and de Dominicis and Hilhorst \cite{deDH} demonstrated a similar connection between the distribution of the cluster weights in the GREM and the cluster weights in the SK model. Motivated by this connection with the SK model, in a seminal paper \cite{Ruelle}, Ruelle gave an alternative, much more explicit and illuminating, description of the Gibbs measure of the GREM in the infinite-volume limit  in terms of a certain family of Poisson processes, as follows. 

\begin{figure}[t]
\centering
\psfrag{a_0}{$\emptyset$}\psfrag{a_1}{$n_1$}\psfrag{a_2}{$(n_1,\ldots,n_{r-1})$}\psfrag{AcodeR}{$\alpha = (n_1,\ldots,n_{r})$}\psfrag{n_code1}{$\Natural^{1}$}\psfrag{n_r1}{$\Natural^{r-1}$}\psfrag{n_r}{$\Natural^r$}
\includegraphics[width=0.7\textwidth]{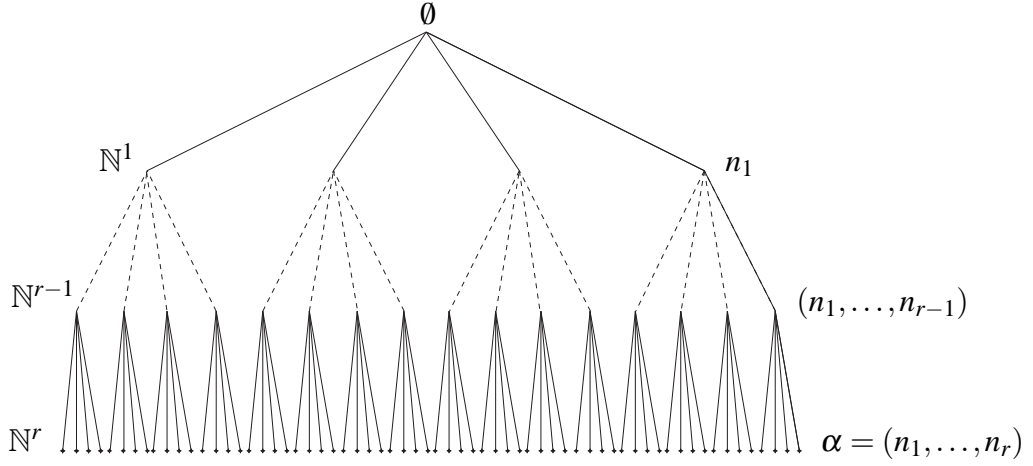}
\caption{\label{Fig1} The leaves $\alpha\in \Natural^r$ index the pure states. The rightmost path is an example of $p(\alpha)$ in (\ref{ch43pathtoleaf}) for one leaf $\alpha$. The figure corresponds to what is called ``$r$-step replica symmetry breaking" in the Parisi ansatz.}
\end{figure}

\paragraph{The Ruelle probability cascades.}
The points and weights of these measures will be indexed by $\Natural^r$ for some fixed $r\geq 1.$ It will be very convenient to think of $\Natural^r$ as the set of leaves of a rooted tree (see Fig. \ref{Fig1}) with the vertex set
\begin{equation}
\A = \Natural^0 \cup \Natural \cup \Natural^2 \cup \ldots \cup \Natural^r,
\label{ch43Atree}
\end{equation}
where $\Natural^0 = \{\emptyset\}$, $\emptyset$ is the root of the tree and each vertex $\alpha=(n_1,\ldots,n_p)\in \Natural^{p}$ for $p\leq r-1$ has children 
$$
\alpha n : = (n_1,\ldots,n_p,n) \in \Natural^{p+1}
$$
for all $n\in \Natural$. Therefore, each vertex $\alpha$ is connected to the root $\emptyset$ by the path
$$
\emptyset \to n_1 \to (n_1,n_2) \to\cdots\to (n_1,\ldots,n_p) = \alpha.
$$
We will denote all the vertices in this path by (the root is not included)
\begin{equation}
p(\alpha) = \bigl\{  n_1, (n_1,n_2),\ldots,(n_1,\ldots,n_p)  \bigr\}.
\label{ch43pathtoleaf}
\end{equation}
The identification of the index set $\Natural^r$ with the leaves of this infinitary tree is very important, because, even though the points in the support of the random measure will be indexed by $\alpha\in \Natural^r$, the construction itself will involve random variables indexed by vertices of the entire tree. For each vertex $\alpha\in \A$, let us denote by $|\alpha|$ its distance from the root of the tree $\emptyset$, or, equivalently, the number of coordinates in $\alpha$, i.e. $\alpha\in \Natural^{|\alpha|}.$ If we recall the parameters (\ref{ch31zetas}) then, for each $\alpha\in \A\setminus \Natural^r$, let $\varPi_\alpha$ be a Poisson process on $(0,\infty)$ with the mean measure 
\begin{equation}
\zeta_{|\alpha|} x^{-1-\zeta_{|\alpha|}} \smsp dx
\label{muzeta}
\end{equation}
and let us generate these processes independently for all such $\alpha$. Let us recall that each Poisson process $\varPi_\alpha$ can be generated by partitioning $(0,\infty)=\cup_{m\geq 1}S_m$ into disjoint sets $S_1 = [1,\infty)$ and $S_m = [1/m,1/(m-1))$ for $m\geq 2$ and then on each set $S_m$ generating independently a Poisson number of points with the mean
$$
\int_{S_m} \! \zeta_{|\alpha|} x^{-1-\zeta_{|\alpha|}}\smsp dx
$$
from the probability distribution on $S_m$ proportional to (\ref{muzeta}). Let us mention that, for technical reasons, it is important that the parameters $\zeta_{|\alpha|}$ in this construction are strictly between $0$ and $1$, which is why we assumed that the inequalities in (\ref{ch31zetas}) are strict. One can arrange all the points in $\varPi_\alpha$ in the decreasing order,
\begin{equation}
u_{\alpha 1} > u_{ \alpha 2} >\ldots >u_{\alpha n} > \ldots,
\label{ch43us}
\end{equation}
and enumerate them using the children $(\alpha n)_{n\geq 1}$ of the vertex $\alpha$. In other words, parent vertices $\alpha \in \A \setminus \Natural^r$ enumerate independent Poisson processes $\varPi_\alpha$ and child vertices $\alpha n\in \A\setminus \Natural^0$ enumerate individual points $u_{\alpha n}$. Given a vertex $\alpha\in \A\setminus \Natural^0$ and the path $p(\alpha)$ in (\ref{ch43pathtoleaf}), we define
\begin{equation}
w_\alpha = \prod_{\beta \in p(\alpha)} u_{\beta}.
\label{ch43ws}
\end{equation}
Finally, for the leaf vertices $\alpha \in \Natural^r$ we define
\begin{equation}
v_\alpha = \frac{w_\alpha}{\sum_{\beta\in \Natural^r} w_\beta}.
\label{ch43vs}
\end{equation}
One can show that the denominator is finite with probability one, so this sequence is well defined. Now, let $e_\alpha$ for $\alpha \in \A\setminus \Natural^0$ be some sequence of orthonormal vectors in $H$. Given this sequence, we consider a set of points $h_\alpha\in H$ indexed by $\alpha\in \Natural^r$,
\begin{equation}
h_\alpha = \sum_{\beta\in p(\alpha)} e_\beta \bigl(q_{|\beta|} - q_{|\beta|-1} \bigr)^{1/2},
\label{ch43hs}
\end{equation}
where the parameters $(q_p)_{0\leq p\leq r}$ were introduced in (\ref{ch31qs}). In other words, as we walk along the path $p(\alpha)$ to the leaf $\alpha\in\Natural^r$, at each step $\beta$ we add a vector in the new orthogonal direction $e_\beta$ of length $\sqrt{q_{|\beta|} - q_{|\beta|-1}}$. We define a random measure $G$ on the Hilbert space $H$ by
\begin{equation}
G(h_\alpha) = v_\alpha 
\,\mbox{ for }\,
\alpha\in \Natural^r.
\label{ch43RPC}
\end{equation}
The measure $G$ is called the {Ruelle probability cascades} (RPC) associated to the parameters (\ref{ch31zetas}) and (\ref{ch31qs}). From the definition (\ref{ch43hs}), it is clear that the scalar product $h_\alpha \cdot h_\beta$ between any two points in the support of $G$  depends only on the number 
\begin{equation}
\alpha\wedge\beta 
:=
 |p(\alpha) \cap p(\beta)  |
\label{ch43wedge}
\end{equation}
of common vertices in the paths from the root $\emptyset$ to the leaves $\alpha, \beta\in \Natural^r$. With this notation, (\ref{ch43hs}) implies that $h_\alpha \cdot h_\beta = q_{\alpha\wedge\beta}.$ Now, if we take three leaves $\alpha,\beta,\gamma \in\Natural^r$ then their paths satisfy
$$
\beta\wedge \gamma \geq \min\bigl(\alpha\wedge \beta, \alpha\wedge \gamma\bigr),
$$
since the vertices shared by the path $p(\alpha)$ with both paths $p(\beta)$ and $p(\gamma)$ will also be shared by $p(\beta)$ and $p(\gamma)$ and, therefore,
\begin{equation}
h_\beta\cdot h_\gamma \geq \min\bigl( h_\alpha\cdot h_\beta, h_\alpha\cdot h_\gamma \bigr),
\label{ch43ultra}
\end{equation}
so the support of $G$ is ultrametric in $H$ by construction. In the work of Ruelle, \cite{Ruelle}, it was stated as an almost evident fact that the Gibbs measure in the Derrida GREM looks like the measure (\ref{ch43RPC}) in the infinite-volume limit, but a detailed proof of this was given later by Bovier and Kurkova in \cite{Bovier}. Because of the connection to the SK model mentioned above, the Ruelle probability cascades are precisely the measures that were expected to describe the Gibbs measures in the SK model in the sense that, asymptotically, the overlap array (\ref{ch12overlaps}) can be approximated in distribution by an overlap array generated by some RPC. The points $(h_\alpha)_{\alpha\in \Natural^r}$ are the pure states and the tree $\A$ can be viewed as a branching tree that indexes the clusters around all the pure states. One can show that the distribution (\ref{Gfop}) of the overlap of two replicas sampled from the Ruelle probability cascades is equal to the distribution function in (\ref{ch30zetafop}), which agrees with the Parisi interpretation of the functional order parameter $\zeta$.  

Such an explicit description of the expected asymptotic Gibbs measures was a very big step, because one could now study their properties using the entire arsenal of the theory of Poisson processes (\cite{king}). Some important properties of the Ruelle probability cascades were already described in the original paper of Ruelle \cite{Ruelle}, while other important properties, which express certain invariance features of these measures, were discovered later by  Bolthausen and Sznitman in \cite{Bolthausen}. We will mention this again below when we talk about the unified stability property in the SK model. 
In the next few section we will explain that all the predictions of the physicists about the structure of the Gibbs measure in the SK and mixed $p$-spin models are, essentially, correct. In general, they hold under a small perturbation of the Hamiltonian, which does not affect the free energy in the infinite-volume limit, but for a class of the so-called generic mixed $p$-spin models they hold precisely, without any perturbation. First, we will explain the connection between the Gibbs measure and the Parisi formula for the free energy.

\section{Free energy and Gibbs measure}

\paragraph{The Aizenman-Sims-Starr scheme.}
Before we describe rigorous results about the structure of the Gibbs measure, let us explain how this structure implies the Parisi formula for the free energy (\ref{ch30Parisi}). For simplicity of notation, we will focus on the Sherrington-Kirkpatrick model (\ref{SKH}) instead of the general mixed $p$-spin model (\ref{mixedH}). We begin with the so-called {Aizenman-Sims-Starr cavity computation}, which was introduced in \cite{AS2}. Let us recall the definition of the partition function $Z_N$ in (\ref{FEZN}) and for $j\geq 0$ let us denote 
\begin{equation}
A_j = \e \log Z_{j+1} - \e \log Z_{j},
\end{equation}
with the convention that $Z_0 = 1$. Then we can rewrite the free energy as follows,
\begin{equation}
F_N = \frac{1}{N}\smsp \e \log Z_N
= \frac{1}{N} \sum_{j=0}^{N-1} A_j.
\label{ch12AS2liminf}
\end{equation}
Clearly, this representation implies that if the sequence $A_N$ converges then its limit is also the limit of the free energy $F_N$. Unfortunately, it is usually difficult to prove that the limit of $A_N$ exists (we will mention one such result at the end, when we talk about generic mixed $p$-spin models) and, therefore, this representation is used only to obtain a lower bound on the free energy, 
\begin{equation}
\liminf_{N\to\infty} F_N \geq \liminf_{N\to\infty} A_N.
\label{FNAN}
\end{equation}
Let us compare the partition functions $Z_N$ and $Z_{N+1}$ and see what they have in common and what makes them different. If we denote $\rho = (\sigma,\eps)\in \varSigma_{N+1}$ for $\sigma\in\varSigma_N$ and $\eps\in\{-1,+1\}$ then we can write
\begin{equation}
H_{N+1}(\rho) = H_N'(\sigma) + \eps z_N(\sigma),
\label{ch12decomp1}
\end{equation}
where
\begin{equation}
H_N'(\sigma) = \frac{1}{\sqrt{N+1}} \sum_{i,j =1}^N g_{ij}\sigma_i \sigma_j
\label{ch12commonH}
\end{equation}
and
\begin{equation}
z_N(\sigma) =  \frac{1}{\sqrt{N+1}} \sum_{i=1}^N \bigl(g_{i(N+1)} + g_{(N+1)i} \bigr)\sigma_i. 
\end{equation}
One the other hand, the part (\ref{ch12commonH}) of the Hamiltonian $H_{N+1}(\rho)$ is, in some sense, also a part of the Hamiltonian $H_N(\sigma)$ since, in distribution, the Gaussian process $H_N(\sigma)$ can be decomposed into a sum of two independent Gaussian processes
\begin{equation}
H_N(\sigma) \stackrel{d}{=}
H_N'(\sigma) + y_N(\sigma),
\label{ch12commonH2}
\end{equation}
where
\begin{equation}
y_N(\sigma) = 
 \frac{1}{\sqrt{N(N+1)}} \sum_{i,j =1}^N g_{ij}'\sigma_i \sigma_j
\end{equation}
for some independent array $(g_{ij}')$ of standard Gaussian random variables. Using the above decompositions (\ref{ch12decomp1}) and (\ref{ch12commonH2}), we can write
\begin{equation}
\e\log Z_{N+1} 
=
\e \log \sum_{\sigma\in\varSigma_N} 2\smsp \ch \bigl(-\beta z_N(\sigma) \bigr) \exp\bigl(- \beta H_{N}'(\sigma)\bigr)
\label{ch12ZN1}
\end{equation}
and
\begin{equation}
\e\log Z_{N} 
=
\e \log \sum_{\sigma\in\varSigma_N} \exp\bigl(-\beta y_N(\sigma) \bigr) \exp \bigl(-\beta H_{N}'(\sigma) \bigr).
\label{ch12ZN}
\end{equation}
Finally, if we consider the Gibbs measure on $\varSigma_N$ corresponding to the Hamiltonian $H_N'(\sigma)$ in (\ref{ch12commonH}),
\begin{equation}
G_N'(\sigma) = \frac{\exp(- \beta H_N'(\sigma))}{Z_N'}
\,\mbox{ where }\,
Z_N' = \sum_{\sigma\in\varSigma_{N}} \exp \bigl( -\beta H_N'(\sigma)\bigr),
\label{ch12MeasureGNprime}
\end{equation}
then (\ref{ch12ZN1}), (\ref{ch12ZN}) can be combined to give the {Aizenman-Sims-Starr representation},
\begin{eqnarray}
A_N
&=&
\e \log \sum_{\sigma\in\Sigma_{N}} 2\smsp \ch\bigl(-\beta z_N(\sigma)\bigr) G_N'(\sigma)
\nonumber
\\
&&
-
\e \log \sum_{\sigma\in\varSigma_{N}} \exp\bigl(-\beta y_N(\sigma)\bigr) G_N'(\sigma).
\label{ch12AS2repr}
\end{eqnarray}
Notice that the Gaussian processes $(z_N(\sigma))$ and $(y_N(\sigma))$ are independent of the randomness of the measure $G_N'$ and have the covariance
\begin{equation}
\e z_N(\sigma^1) z_N(\sigma^2) = 2 R_{1,2} + O(N^{-1}),\,\,
\e y_N(\sigma^1) y_N(\sigma^2) = R_{1,2}^2 + O(N^{-1}).
\label{ch12Covzy}
\end{equation}
Suppose that we replace the Gibbs measure $G_N'$ in (\ref{ch12AS2repr}) by the Ruelle probability cascades $G$ in (\ref{ch43RPC}) and replace the Gaussian processes $(z_N(\sigma))$ and $(y_N(\sigma))$ by Gaussian processes $(z(h_\alpha))$ and $(y(h_\alpha))$ indexed by the points $(h_\alpha)_{\alpha\in \Natural^r}$ in the support of $G$ with the same covariance structure as (\ref{ch12Covzy}),
\begin{equation}
\e z(h_{\alpha}) z(h_{\beta}) = 2 \smsp h_{\alpha}\cdot h_{\beta},\,\,
\e y(h_{\alpha}) y(h_{\beta}) = (h_{\alpha}\cdot h_{\beta})^2.
\label{ch12CovzyG}
\end{equation}
Such processes are very easy to construct explicitly, if we recall the definition of the points $h_\alpha$ in (\ref{ch43hs}). Namely, let $(\eta_\alpha)_{\alpha \in \A\setminus \Natural^0}$ be a sequence of i.i.d. standard Gaussian random variables and, for each $p\geq 1$, let us define a family of Gaussian random variables indexed by $(h_\alpha)_{\alpha\in \Natural^r}$,
\begin{equation}
g_p(h_\alpha) = \sum_{\beta\in p(\alpha)} \eta_\beta (q_{|\beta|}^p - q_{|\beta|-1}^p)^{1/2}.
\label{ch43etas}
\end{equation}
Recalling the notation (\ref{ch43wedge}), it is obvious that the covariance of this process is 
\begin{equation}
\e g_p(h_\alpha)g_p(h_\beta) = q_{\alpha\wedge \beta}^p = (h_\alpha \cdot h_\beta)^p,
\label{ch43Cov}
\end{equation}
so we can take $z(h_\alpha) = \sqrt{2} g_1(h_\alpha)$ and $y(h_\alpha) = g_2(h_\alpha)$. Then the functional (\ref{ch12AS2repr}) will be replaced by
\begin{equation}
\PP(\zeta) =
\e \log \sum_{\alpha\in \Natural^r}  2 \smsp \ch \bigl(-\beta z(h_\alpha)\bigr) v_\alpha
-
\e \log \sum_{\alpha\in \Natural^r}  \exp \bigl( -\beta y(h_\alpha)\bigr) v_\alpha.
\label{ch30PzetaRPC}
\end{equation}
Writing $\PP(\zeta)$ here is not an abuse of notation, since it turns out that the right hand side coincides with the Parisi functional in (\ref{ch30Pzeta}) when $\xi(x) = x^2$ and $\theta(x) = x^2$, which is precisely the case of the Sherrington-Kirkpatrick model. The equality of these two different representations can be proved using the properties of the Poisson processes with the mean measures (\ref{muzeta}) that appear in the definition of the Ruelle probability cascades, and (\ref{ch30PzetaRPC}) gives a very natural interpretation of the Parisi functional $\PP(\zeta)$ in (\ref{ch30Pzeta}). It remains to explain that, if we assume the Parisi ansatz for the Gibbs measure, then the connection between (\ref{ch12AS2repr}) and (\ref{ch30PzetaRPC}) is more than just a formal resemblance and that together with (\ref{FNAN}) it implies that 
\begin{equation}
\liminf_{N\to\infty} F_N \geq \inf_{\zeta} \PP(\zeta).
\label{FNPzeta}
\end{equation}
This is again a consequence of the fundamental fact that we mentioned above, namely, that all the relevant information about the Gibbs measure in the SK model is contained in the overlap matrix. In the present context, it is not difficult to show that, due to the covariance structure of the Gaussian processes (\ref{ch12Covzy}) and (\ref{ch12CovzyG}), the quantities $A_N$ in (\ref{ch12AS2repr}) and $\PP(\zeta)$ in (\ref{ch12CovzyG}) are, in fact, given by the same continuous functional of the distribution of the overlap arrays $R^N$ and $R$ generated by i.i.d. samples of replicas from $G_N'$ and $G$ correspondingly. As a result, if we consider a subsequence along which the $\liminf_{N\to\infty} A_N$ is achieved and, at the same time, the array $R^N$ converges in distribution to some array $R^\infty$, then the lower limit of $A_N$ can be written as the same functional of the distribution of $R^\infty$. Finally, if we believe that the predictions of the physicists are correct, we can approximate $R^\infty$ in distribution by the overlap arrays $R$ generated by the Ruelle probability cascades and, therefore, the lower limit is bounded from below by $\inf_{\zeta} \PP(\zeta)$, which proves (\ref{FNPzeta}). The main difficulty in this approach is to show that the Parisi ansatz for the Gibbs measure and the overlap array is, indeed, correct in the infinite-volume limit, which will be discussed below.

\paragraph{Guerra's replica symmetry breaking bound.}
The fact that the Parisi formula also gives an upper bound on the free energy,
\begin{equation}
\limsup_{N\to\infty} F_N \leq \inf_{\zeta} \PP(\zeta),
\label{FNPzeta2}
\end{equation}
was proved in a breakthrough work of Guerra, \cite{Guerra}. The original argument in \cite{Guerra} was given in the language of the recursive formula (\ref{ch30Xl}), but, as was observed in \cite{AS2}, it can also be written in the language of the Ruelle probability cascades. The essence of Guerra's result is the following interpolation between the SK model and the Ruelle probability cascades. Let  $(z_i(h_\alpha))$ and $(y_i(h_\alpha))$ for $i\geq 1$ be independent copies of the processes $(z(h_\alpha))$ and $(y(h_\alpha))$ in (\ref{ch12CovzyG}) and, for $0\leq t\leq 1$, let us consider the Hamiltonian
\begin{equation}
H_{N,t}(\sigma,h_\alpha) = 
\sqrt{t} H_N(\sigma) + \sqrt{1-t}\smsp \sum_{i=1}^N z_{i}(h_\alpha) \sigma_i 
+\sqrt{t}\smsp  \sum_{i=1}^N y_{i}(h_\alpha)
\label{ch33Hta}
\end{equation}
indexed by vectors $(\sigma,h_\alpha)$ such that $\sigma$ belongs to the support $\Sigma_N$ of the Gibbs measure $G_N$ and $h_\alpha$ belongs to the support of the measure $G$ in (\ref{ch43RPC}). To this Hamiltonian one can associate the free energy
\begin{equation}
\varphi(t)=\frac{1}{N}\smsp \e\log \sum_{\sigma,\alpha} v_{\alpha} \exp \bigl( -\beta H_{N, t}(\sigma,h_\alpha) \bigr)
\label{ch33Gint}
\end{equation}
and, by a straightforward computation using the Gaussian integration by parts, one can check that $\varphi'(t)\leq 0$ and, therefore, $\varphi(1)\leq \varphi(0).$  It is easy to see that
$$
\varphi(0)
=
\frac{1}{N}\smsp \e\log \sum_{\alpha\in\Natural^r} v_{\alpha} \prod_{i\leq N}
2\ch\bigl( -\beta z_{i}(h_\alpha) \bigr)
$$
and
$$
\varphi(1)
=
F_N
+
\frac{1}{N}\smsp \e\log \sum_{\alpha\in\Natural^r} v_{\alpha} \prod_{i\leq N} 
\exp \bigl(-\beta y_{i}(h_\alpha) \bigr).
$$
It is, again, a consequence of the properties of the Poisson processes involved in the construction of the Ruelle probability cascades that, in fact, the independent copies for $i\leq N$ can be decoupled here and
$$
\varphi(0)
=
 \e\log \sum_{\alpha\in\Natural^r}  2\smsp \ch\bigl(-\beta z(h_\alpha)\bigr) v_{\alpha}
$$
and
$$
\varphi(1)
=
F_N
+
\e\log \sum_{\alpha\in\Natural^r} \exp\bigl(-\beta y(h_\alpha)\bigr) v_{\alpha} .
$$
Recalling the representation (\ref{ch30PzetaRPC}), the inequality $\varphi(1)\leq \varphi(0)$ can be written as $F_N\leq \PP(\zeta)$, which yields the upper bound (\ref{FNPzeta2}). After Guerra's discovery of the above interpolation argument, Talagrand proved in his famous tour-de-force paper \cite{TPF} that the Parisi formula, indeed, gives the free energy in the SK model in the infinite-volume limit. Talagrand's ingenious proof finds a way around the Parisi ansatz for the Gibbs measure, but it is rather involved. The Aizenman-Sims-Starr scheme above gives a more natural approach if we are able to confirm the Parisi ansatz for the asymptotic Gibbs measures. Moreover, the argument in \cite{TPF} works only for mixed $p$-spin models for even $p\geq 2$, while the above approach can be modified to yield the Parisi formula in the case when odd $p$-spin interactions are present as well (see \cite{PPF}). Nevertheless, to understand the impact of the results of Guerra \cite{Guerra} and Talagrand \cite{TPF} (proved in 2003), one only needs to remember that a proof of the existence of the limit of the free energy by Guerra and Toninelli in \cite{GuerraToninelli} was quite an impressive result only a year earlier.

\section{Stability of the Gibbs measure}

\paragraph{The Ghirlanda-Guerra identities.}
Below we will explain an approach to proving the predictions of the physicists for the Gibbs measure based on the so called Ghirlanda-Guerra identities. These identities were first discovered by Ghirlanda and Guerra in \cite{GG} in the setting of the mixed $p$-spin models, where they were proved on average over the parameters $(\beta_p)$ in (\ref{mixedH}). However, the general idea can be used in many other models if we utilize the mixed $p$-spin Hamiltonian in the role of a perturbation. We will try to emphasize wide applicability of this idea by giving some mild sufficient conditions that ensure the validity of these identities. For all $p\geq 1$, let us consider
\begin{equation}
g_{p}(\sigma)
=
\frac{1}{N^{p/2}}
\sum_{i_1,\ldots,i_p = 1}^N g_{i_1\ldots i_p}' \sigma_{i_1}\ldots\sigma_{i_p},
\label{ch31mixedppert}
\end{equation}
where the random variables $(g_{i_1\ldots i_p}')$ are i.i.d. standard Gaussian and independent of everything else, and define
\begin{equation}
g(\sigma) = \sum_{p\geq 1} 2^{-p} x_p\smsp g_{p}(\sigma)
\label{ch31mixedHpert}
\end{equation}
for some parameters $(x_p)_{p\geq 1}$ that belong to the interval $x_p\in[0,3]$ for all $p\geq 1$. This Gaussian process is of the same nature as the mixed $p$-spin Hamiltonian (\ref{mixedH}) except for a different normalization in (\ref{ch31mixedppert}), which implies that the covariance
\begin{equation}
\e g(\sigma^1) g(\sigma^2) = \sum_{p\geq 1} 4^{-p} x_p^2\smsp R_{1,2}^p.
\label{ch31Covxipert}
\end{equation}
In other words, $g(\sigma)$ is of a smaller order than $H_N(\sigma)$ because of the additional factor $N^{-1/2}$. Let us now consider a model with an arbitrary Hamiltonian $H(\sigma)$ on $\varSigma_N$, either random or non-random, and consider the perturbed Hamiltonian
\begin{equation}
H^{\mathrm{pert}}(\sigma) = H(\sigma) + s g(\sigma),
\label{ch31Hpert}
\end{equation} 
for some parameter $s\geq 0.$ What is the advantage of adding the perturbation term (\ref{ch31mixedHpert}) to the original Hamiltonian of the model? The answer to this question lies in the fact that, under certain conditions, this perturbation term, in some sense, regularizes the Gibbs measure and forces it to satisfy useful properties without affecting our mail goal---the computation of the free energy. Using (\ref{ch31Covxipert}) and the independence of $g(\sigma)$ and $H(\sigma)$, it is easy to see that
\begin{eqnarray}
\frac{1}{N}\smsp \e\log \sum_{\sigma\in\varSigma_N} \exp  H(\sigma)
& \leq &
\frac{1}{N}\smsp \e\log \sum_{\sigma\in\varSigma_N} \exp  \bigl(H(\sigma) + s g(\sigma)\bigr)
\label{ch31FEsamepert}
\\
& \leq &
\frac{1}{N}\smsp \e\log \sum_{\sigma\in\varSigma_N} \exp  H(\sigma)
+ \frac{s^2}{2N} \sum_{p\geq 1} 4^{-p} x_p^2.
\nonumber
\end{eqnarray}
Both inequalities follow from Jensen's inequality applied either to the sum or the expectation with respect to $g(\sigma)$ conditionally on $H(\sigma)$. This implies that if we let $s=s_N$ in (\ref{ch31Hpert}) depend on $N$ in such a way that
\begin{equation}
\lim_{N\to\infty} N^{-1} s_N^2 = 0, 
\label{ch31tN}
\end{equation} 
then the limit of the free energy is unchanged by the perturbation term $s g(\sigma).$ On the other hand, if $s=s_N$ is not too small then it turns out that the perturbation term has a non-trivial influence on the Gibbs measure of the model. Consider a  function
\begin{equation}
\varphi =
\log \sum_{\sigma\in\varSigma_N} \exp \bigl(H(\sigma) + s g(\sigma)\bigr)
\label{ch31theta}
\end{equation}
that will be viewed as a random function $\varphi = \varphi\bigl((x_p)\bigr)$ of the parameters $(x_p)$, and suppose that
\begin{equation}
\sup\Bigl\{ \e |\varphi - \e \varphi | : 0\leq x_p\leq 3, p\geq 1\Bigr\}\leq v_N(s)
\label{ch31vt}
\end{equation}
for some function $v_N(s)$ that describes how well $\varphi\bigl((x_p)\bigr)$ is concentrated around its expected value uniformly over all possible choices of the parameters $(x_p)$ from the interval $[0,3].$ Main condition about the model will be expressed in terms of this concentration function, namely, that there exists a sequence $s= s_N$ such that
\begin{equation}
\lim_{N\to\infty} s_N=\infty
\,\mbox{ and }\,
\lim_{N\to\infty} s_N^{-2} v_N(s_N) = 0.
\label{ch31GGassumption}
\end{equation}
Of course, this condition will be useful only if the sequence $s_N$  also satisfies (\ref{ch31tN}) and the perturbation term does not affect the limit of the free energy. In the case of the mixed $p$-spin model, $H(\sigma) = -\beta H_N(\sigma)$ for $H_N(\sigma)$ defined in (\ref{mixedH}), one can easily check using some standard Gaussian concentration inequalities that (\ref{ch31tN}) and (\ref{ch31GGassumption}) hold with the choice of $s_N = N^{\gamma}$ for any $1/4<\gamma< 1/2$. One can also check that this condition holds in other models, for example, random $p$-spin and $K$-sat models, or for any non-random Hamiltonian $H(\sigma)$. Now, let 
\begin{equation}
G_N(\sigma) = \frac{\exp  H^{\mathrm{pert}}(\sigma)}{Z_N},
\,\mbox{ where }\,
Z_N = \sum_{\sigma\in\varSigma_{N}} \exp  H^{\mathrm{pert}}(\sigma),
\label{ch31GNpert}
\end{equation}
be the Gibbs measure corresponding to the perturbed Hamiltonian (\ref{ch31Hpert}) and let $\la\cdot \ra$ denote the average with respect to $G_N^{\otimes \infty}$, or all replicas. For any $n\geq 2, p\geq 1$ and any function $f$ of the overlaps $(R_{l,l'})_{ l,l'\leq n}$ of $n$ replicas, let us define
\begin{equation}
\varDelta(f,n,p) = 
\Bigl|
\e  \bigl\la f R_{1,n+1}^p \bigr\ra -  \frac{1}{n}\e \bigl\la f \bigr\ra \smsp \e\bigl\la R_{1,2}^p\bigr\ra - \frac{1}{n}\sum_{l=2}^{n}\e \bigl\la f R_{1,l}^p\bigr\ra
\Bigr|.
\label{ch31GG}
\end{equation}
If we now think of the parameters $(x_p)_{p\geq 1}$ in (\ref{ch31mixedHpert}) as a sequence of i.i.d. random variables with the uniform distribution on $[1,2]$ and denote by $\e_x$ the expectation with respect to such sequence then (\ref{ch31GGassumption}) is a sufficient condition to guarantee that 
\begin{equation}
\lim_{N\to\infty} \e_x \smsp \varDelta(f,n,p) = 0.
\label{ch31GGxlim}
\end{equation} 
Once we have this statement on average, we can, of course, make a specific non-random choice of parameters $(x_p^N)_{p\geq 1}$, which may vary with $N$, such that 
\begin{equation}
\lim_{N\to\infty} \smsp \varDelta(f,n,p) = 0.
\label{ch31GGxlim2}
\end{equation} 
 In the thermodynamic limit, this can be expressed as a property of the asymptotic Gibbs measures. Let us consider any subsequential limit of the distribution of the overlap matrix $R^N$ generated by the perturbed Gibbs measure $G_N$ in (\ref{ch31GNpert}) and let $G$ be the corresponding asymptotic Gibbs measure on a Hilbert space defined via the Dovbysh-Sudakov representation. If we still denote by $\la\cdot \ra$ the average with respect to $G^{\otimes \infty}$ then (\ref{ch31GGxlim2}) implies the Ghirlanda-Guerra identities,
\begin{equation}
\e  \bigl\la f R_{1,n+1}^p \bigr\ra 
=
 \frac{1}{n}\e \bigl\la f \bigr\ra \smsp \e\bigl\la R_{1,2}^p\bigr\ra 
 +
 \frac{1}{n}\sum_{l=2}^{n}\e \bigl\la f R_{1,l}^p\bigr\ra,
\label{ch31GGlimit}
\end{equation}
for any $n\geq 2, p\geq 1$ and any function $f$ of the overlaps $(R_{l,l'})_{ l,l'\leq n}=(\sigma^l\cdot\sigma^{l'})_{ l,l'\leq n}$ of $n$ replicas sampled from $G.$ Of course, since any bounded measurable function $\psi$ can be approximated by polynomials (in the $L^1$ sense), 
\begin{equation}
\e  \bigl\la f \psi(R_{1,n+1}) \bigr\ra 
=
 \frac{1}{n}\e \bigl\la f \bigr\ra \smsp \e\bigl\la \psi(R_{1,2}) \bigr\ra 
 +
 \frac{1}{n}\sum_{l=2}^{n}\e \bigl\la f \psi(R_{1,l})\bigr\ra.
\label{ch31GGlimit2}
\end{equation}
If $\zeta$ is the distribution of one overlap, as in (\ref{Gfop}), then (\ref{ch31GGlimit2}) can be expressed by saying that, conditionally on $(R_{l,l'})_{l,l' \leq n},$ the distribution of $R_{1,n+1}$ is given by the mixture 
$$
n^{-1} \zeta + n^{-1} \sum_{l=2}^n \delta_{R_{1,l}}.
$$ 
These identities already appear in the Parisi replica method where they arise as a consequence of ``replica equivalence", but Ghirlanda and Guerra gave the first mathematical proof using the self-averaging of the free energy, which is what the condition (\ref{ch31GGassumption}) basically means. The self-averaging of the free energy in the Sherrington-Kirkpatrick model was first proved by Pastur and Shcherbina in \cite{Pastur}. The identities (\ref{ch31GGlimit}) might look mysterious, but, in fact, they are just a manifestation of the general principle of the concentration of a Hamiltonian, in this case
\begin{equation}
\lim_{N\to\infty} \e_x \e \bigl\la \bigl|g_p(\sigma) -\e\bigl\la g_p(\sigma)\bigr\ra \bigr|\bigr\ra =0, 
\label{ch31GGbasic}
\end{equation}
which can be proved using the self-averaging of the free energy condition (\ref{ch31GGassumption}). The way (\ref{ch31GGbasic}) implies the Ghirlanda-Guerra identities is very simple, essentially, by testing this concentration on a test function. If we fix $n\geq 2$ and consider a bounded function $f=f((R_{l,l'})_{ l,l'\leq n})$ of the overlaps of $n$ replicas then
\begin{equation}
\bigl|\e\bigl\la f g_p(\sigma^1) \bigr\ra 
- \e \bigl \la f \bigr\ra \e \bigl\la g_p(\sigma) \bigr\ra \bigr|
\leq
\|f\|_\infty
\e \bigl\la \bigl| g_p(\sigma) - \e \bigl\la g_p(\sigma) \bigr\ra\bigr|\bigr\ra
\label{ch31GGintegrate}
\end{equation}
and (\ref{ch31GGbasic}) implies that
\begin{equation}
\lim_{N\to\infty} \e_x \bigl|\e\bigl\la f g_p(\sigma^1) \bigr\ra 
- \e \bigl \la f \bigr\ra \e \bigl\la g_p(\sigma) \bigr\ra \bigr| =0. 
\label{ch31GGbasic2}
\end{equation}
This is precisely the equation (\ref{ch31GGxlim}) after we use the Gaussian integration by parts. 

\paragraph{The Aizenman-Contucci stochastic stability}
Another famous property of the Gibbs measure, the so-called stochastic stability discovered by  Aizenman and Contucci in \cite{AC}, is also a consequence of (\ref{ch31GGbasic}). A proof can be found in \cite{Tal-New} (see also \cite{CGSS}) and a rigorous justification of how to extend the stochastic stability to the setting of the asymptotic Gibbs measures can be found in \cite{ACh}. To state this property, let us assume, for simplicity, that the asymptotic Gibbs measure $G$ is atomic, $G(h_l)=v_l,$
with the weights arranged in non-increasing order, $v_1\geq v_2\geq \ldots$ . Given integer $p\geq 1,$ let $(g_p(h_l))_{l\geq 1}$ be a Gaussian sequence conditionally on $G$ indexed
by the points $(h_l)_{l\geq 1}$ with the covariance
\begin{equation}
\e g_p(h_l) g_p(h_{l'}) = (h_l\cdot h_{l'})^p,
\label{Covp}
\end{equation}
which is reminiscent of (\ref{ch43Cov}), only now we do not know a priori that the support $\{h_l: l\geq 1\}$ is ultrametric in $H$. Given $t\in\Reals,$ consider a new measure 
\begin{equation}
G_t(h_l) = v_l^t=\frac{v_l \exp t g_p(h_l)}{\sum_{j\geq 1} v_j \exp t g_p(h_j)}
\label{density}
\end{equation}
defined by the random change of density proportional to $\exp tg_p(h_l)$. Then the Aizenman-Contucci stochastic stability, basically, states that this new measure generates the same overlap array in distribution as the original measure $G$. One way to express this property is as follows. Let $\pi:\Natural\to \Natural$ be a permutation such that the weights $(v_{\pi(l)}^t)$ are also arranged in non-increasing order. Then, 
\begin{equation}
\Bigl(
\bigl(v_{\pi(l)}^t \bigr)_{l\geq 1}, \bigl(h_{\pi(l)}\cdot  h_{\pi(l')}\bigr)_{l,l'\geq 1}
\Bigr)
\stackrel{d}{=}
\Bigl(
\bigl(v_l \bigr)_{l\geq 1}, \bigl(h_l\cdot h_{l'} \bigr)_{l,l'\geq 1}
\Bigr)
\label{AC}
\end{equation}
for any $p\geq 1$ and $t\in \Reals$, which, clearly, implies that the overlap arrays generated by these measures will have the same distribution. It is not difficult to see that, in fact, the two statements are equivalent.

\paragraph{Unified stability property.}
Even though the proof of the Parisi ansatz that will be described in the next section is based only on the Ghirlanda-Guerra identities, the reason we mention the Aizenman-Contucci stochastic stability (\ref{AC}) is because, in a number of ways, it played a very important role in the development of the area. In particular, one of the main ideas behind the proof of the Parisi ansatz was first discovered using a unified stability property, proved in \cite{ACGG}, that combines the Ghirlanda-Guerra identities (\ref{ch31GGlimit}) and the Aizenman-Contucci stochastic stability (\ref{AC}). It can be stated with the notation in (\ref{AC}) as follows. It is well known (we will discuss this again below) that if the measure $G$ satisfies the Ghirlanda-Guerra identities and if $c^2$ is the largest point of the support of the distribution of the overlap $R_{1,2}$ under $\e G^{\otimes 2}$ then, with probability one, $G$ is concentrated on the sphere of radius $c$. Let
\begin{equation}
b_p = (c^2)^p - \e\la R_{1,2}^p\ra.
\label{bp}
\end{equation}
Then, a random measure $G$ satisfies the Ghirlanda-Guerra identities (\ref{ch31GGlimit}) and the Aizenman-Contucci stochastic stability (\ref{AC}) if and only if it is concentrated on the sphere of constant radius, say $c$,  and for any $p\geq 1$ and $t\in\Reals,$
\begin{eqnarray}
&&
\Bigl(
\bigl(v_{\pi(l)}^t \bigr)_{l\geq 1}, \bigl(g_p(h_{\pi(l)}) - b_p t\bigr)_{l\geq 1}, \bigl(h_{\pi(l)}\cdot  h_{\pi(l')}\bigr)_{l,l'\geq 1}
\Bigr)
\nonumber
\\
&&\stackrel{d}{=}
\Bigr(
\bigl(v_l\bigr)_{l\geq 1}, \bigl(g_p(h_l)\bigr)_{l\geq 1}, \bigl(h_l\cdot h_{l'} \bigr)_{l,l'\geq 1}
\Bigr).
\label{main}
\end{eqnarray}
Comparing with (\ref{AC}), we see that the Ghirlanda-Guerra identities are now replaced by the statement that, after the permutation $\pi$ which rearranges the weights in (\ref{density}) in the decreasing order, the distribution of the Gaussian process $(g_p(h_l))$ will only be affected by a constant shift $b_p t$. Interestingly, the unified stability property (\ref{main}) was known for some time in the setting of the Ruelle probability cascades, where it was proved by Bolthausen and Sznitman in \cite{Bolthausen} using properties of the Poisson processes in the construction of the Ruelle probability cascades. However, the Ghirlanda-Guerra identities for the RPC were originally proved by Talagrand \cite{SG} and Bovier and Kurkova \cite{Bovier} by analyzing the Gibbs measure in the Derrida REM and GREM, and it was only later noticed by Talagrand that they follow much more easily from the Bolthausen-Sznitman invariance. The main result of \cite{ACGG} stated in (\ref{main}), basically, reverses Talagrand's observation.

\section{Structure of the Gibbs measure}

It was clear since they were discovered that the stability properties impose strong constraints on the structure of the Gibbs measure, but the question was whether they lead all the way to the Ruelle probability cascades? The first partial answer to this question was given in an influential work of Arguin and Aizenman, \cite{AA}, who proved that, under a  technical assumption that the overlap takes only finitely many values in the thermodynamic limit, 
$R_{1,2} \in \{q_0,\ldots, q_r \}$,
the Aizenman-Contucci stochastic stability (\ref{AC}) implies the ultrametricity predicted by the Parisi ansatz. Soon after, it was shown in \cite{PGG} under the same technical assumption that the Ghirlanda-Guerra identities also imply ultrametricity (an elementary proof can be found in \cite{PGG2}). Another approach was given by Talagrand in \cite{Tal-New}. However, since at low temperature the overlap does not necessarily take finitely many values in the thermodynamic limit, all these result were not directly applicable to the SK and mixed $p$-spin models. Nevertheless, they strongly suggested that the stability properties can explain the Parisi ansatz and, indeed, it was recently proved in \cite{PUltra} that the Ghirlanda-Guerra identities imply ultrametricity (\ref{ultra2}) without any technical assumptions. Before we explain some of the main ideas in the proof, let us first describe several preliminary facts that follow from the Ghirlanda-Guerra identities.

\paragraph{Pure states.}
First of all, one can show very easily (see \cite{PGG}) that the Ghirlanda-Guerra identities (\ref{ch31GGlimit}) yield the Parisi pure states picture described above. Namely, if $c^2$ is the largest point in the support of the distribution $\zeta$ of the overlap $R_{1,2}$ under $\e G^{\otimes 2}$ defined in (\ref{Gfop}) then, with probability one, $G(\|h\| = c)=1$. Moreover, the measure $G$ is purely atomic if $\zeta(\{c^2\})>0$, otherwise, it has no atoms. Of course, it is clear that Parisi's pure states picture in \cite{Parisi83} was meant to be understood in approximate sense and, when $\zeta(\{c^2\})=0$ and $G$ has not atoms, we can create pure states using ultrametricity by considering equivalence clusters in (\ref{diameter}) for small positive diameter $d>0$. In the case when $\zeta(\{c^2\})>0$, the pure states picture holds not only for the asymptotic Gibbs measures in the infinite-volume limit, but also for the original Gibbs measures $G_N$ for finite size systems in some approximates sense, as was shown by Talagrand in \cite{Tal-New}.

\paragraph{Talagrand's positivity principle.}
Another important consequence of the Ghirlanda-Guerra identities is the so-called Talagrand positivity principle, proved in \cite{SG}, which states that the overlaps can take only non-negative values in the thermodynamic limit, so $\sigma^1\cdot\sigma^2\geq 0$ for any two points in the support of $G$. In the Parisi replica method, the overlap was always assumed to be non-negative due to the symmetry breaking, and we see that this, indeed, can be obtained using a small perturbation of the Hamiltonian which ensures the validity of the Ghirlanda-Guerra identities. One key application of the positivity principle is to show that Guerra's interpolation argument leading to the upper bound (\ref{FNPzeta2}) also works for mixed $p$-spin models that include the pure $p$-spin Hamiltonians for odd $p\geq 3$, which was observed by Talagrand in \cite{T2}.

\paragraph{Characterizing asymptotic Gibbs' measures.}
Finally, let us mention a fact that has been well known since the discovery of the Ghirlanda-Guerra identities, namely, that together with ultrametricity these identities  determine the distribution of the entire overlap array uniquely in terms of the functional order parameter $\zeta$ and, moreover, one can approximate the overlap array in distribution by an overlap array generated by some Ruelle probability cascades. In other words, as soon as we have ultrametricity, all the predictions of the physicists are confirmed. The idea here is very straightforward and we will only illustrate it in the simplest case when the overlaps take finitely many values, $R_{1,2} \in \{q_0,\ldots, q_r \}$. The general case easily follows by approximation. In the discrete case, we only need to demonstrate that, using ultrametricity and the Ghirlanda-Guerra identities, we can compute in terms of $\zeta$ the probability of any particular configuration of finitely many overlaps, 
\begin{equation}
\e \Bigl\la I\Bigl(R_{l,l'} = q_{l,l'} : l\not = l'\leq n+1\Bigr) \Bigr\ra,
\label{ch44ultraq1}
\end{equation}
for any $n\geq 1$ and any $q_{l,l'}\in \{q_0,\ldots, q_r\}$. Let us find the largest elements among $q_{l,l'}$ for $l\not = l'$ and, without loss of generality, suppose that $q_{1,n+1}$ is one of them. We only have to consider $(q_{l,l'})$ that are ultrametric, since, otherwise, (\ref{ch44ultraq1}) is equal to zero. In particular, since $q_{1,n+1}$ is the largest, for $2\leq l \leq n$,
$$
q_{1,l} \geq \min\bigl(q_{1,n+1}, q_{l,n+1}\bigr) = q_{l,n+1}
$$
and
$$
q_{l,n+1} \geq \min\bigl(q_{1,n+1}, q_{1,l}\bigr) = q_{1,l},
$$  
which implies that $q_{1,l} = q_{l,n+1}$. Hence, if the overlap $R_{1,n+1} = q_{1,n+1}$ then, for all $2\leq l\leq n$, $R_{1,l} = q_{1,l}$ automatically implies that $R_{l,n+1} = q_{1,l}$ and (\ref{ch44ultraq1}) equals
\begin{equation}
\e \bigl\la I\bigl(R_{l,l'} = q_{l,l'} : l,l'\leq n\bigr) I\bigl(R_{1,n+1} = q_{1,n+1}\bigr) \bigr\ra.
\label{ch44ultraq20}
\end{equation}
In other words, if we know that the replicas $\sigma^1$ and $\sigma^{n+1}$ are the closest then, due to ultrametricity, all the conditions $R_{l,n+1} = q_{l,n+1} = q_{1,l}$ become redundant and we can omit them. The quantity (\ref{ch44ultraq20}) is now of the same type as the left hand side of (\ref{ch31GGlimit2}) and, therefore, the Ghirlanda-Guerra identities imply that it is equal to 
\begin{eqnarray}
&&
\frac{1}{n}\zeta\bigl(\{q_{1,n+1}\}\bigr)\e \bigl\la I\bigl(R_{l,l'} = q_{l,l'} : l,l'\leq n\bigr) \bigr\ra
\label{ch44ultraq2}
\\
&&+
\frac{1}{n}\sum_{l=2}^n I\bigl(q_{1,l} = q_{1,n+1}\bigr) \e \bigl\la I\bigl(R_{l,l'} = q_{l,l'} : l,l'\leq n \bigr) \bigr\ra.
\nonumber
\end{eqnarray}
We can continue this computation recursively over $n$ and, in the end, (\ref{ch44ultraq1}) will be expressed completely in terms of the distribution of one overlap, $\zeta$. To conclude that the overlap array can actually be generated by the Ruelle probability cascades corresponding to the functional order parameter $\zeta$, we only need to recall that both properties, the Ghirlanda-Guerra identities and ultrametricity, are satisfied by the RPC, so all the probabilities (\ref{ch44ultraq1}) will be given by the same computation. 

\paragraph{Ultrametricity.}
It remains to explain the main result in \cite{PUltra} which shows that ultrametricity is also a consequence of the Ghirlanda-Guerra identities. The main idea of the proof is the following invariance property. Given $n\geq 1$, we consider $n$ bounded measurable functions $f_1,\ldots, f_n:  \Reals\to\Reals$ and let
\begin{equation}
F(\sigma,\sigma^1,\ldots,\sigma^n) = f_1(\sigma\cdot\sigma^1)+\ldots+f_n(\sigma\cdot\sigma^n).
\label{ch45F1}
\end{equation}
For $1\leq l\leq n$, we define
\begin{equation}
F_l(\sigma,\sigma^1,\ldots,\sigma^n) = F(\sigma,\sigma^1,\ldots,\sigma^n)
 - f_l( \sigma\cdot\sigma^l)+ \e \bigl \la f_l(R_{1,2}) \bigr\ra,
\label{ch45F2}
\end{equation}
where, as before, $\la\cdot \ra$ denotes the average with respect to $G^{\otimes \infty}$. Then, for any bounded measurable function $\varPhi$ of the overlaps $(R_{l,l'})_{l,l'\leq n}$ of $n$ replicas,
\begin{equation}
\e \bigl\la\varPhi \bigr \ra =
\e\Bigl\la
\varPhi\smsp
\frac{\exp \sum_{l=1}^{n} F_l(\sigma^l,\sigma^1,\ldots,\sigma^n)}
{\la\exp F(\sigma,\sigma^1,\ldots,\sigma^n)\ra_{\hspace{-0.3mm}\mathunderscore}^n}
\Bigr\ra,
\label{ch45main}
\end{equation}
where $\la\cdot\ra_{\hspace{-0.3mm}\mathunderscore}$ in the denominator is the average in $\sigma$ for fixed $\sigma^1,\ldots, \sigma^n$ with respect to the measure $G$. One can think of the ratio on the right hand side as a change of density that does not affect the distribution of the overlaps of $n$ replicas. Originally, this invariance property was discovered using the unified stability property (\ref{main}), so the Aizenman-Contucci stochastic stability played an equally important role. 
However, the proof presented in \cite{PUltra} is much more simple and straightforward, and is based only on the
the Ghirlanda-Guerra identities. In some sense, this is good news because the Aizenman-Contucci stability is a more subtle property to work with than the Ghirlanda-Guerra identities, especially in the thermodynamic limit. To prove (\ref{ch45main}), one can consider an interpolating function
\begin{equation}
\varphi(t) = 
\e\Bigl\la
\varPhi
\frac{\exp \sum_{l=1}^{n} t F_l(\sigma^l,\sigma^1,\ldots,\sigma^n)}
{\la\exp t F(\sigma,\sigma^1,\ldots,\sigma^n)\ra_{\hspace{-0.3mm}\mathunderscore}^n}
\Bigr\ra
\label{ch45varphitdefine}
\end{equation}
and, using an elementary calculation, check that the Ghirlanda-Guerra identities imply that all the derivatives vanish at zero, $\varphi^{(k)}(0)=0$. Taylor's expansion and some basic estimates of the derivatives yield that this function is constant for all $t\geq 0$, proving that $\varphi(0)=\varphi(1)$, which is precisely (\ref{ch45main}).  A special feature of the invariance property (\ref{ch45main}) is that it contains some very useful information not only about the overlaps but also about the Gibbs weights of the neighborhoods of the replicas $\sigma^1,\ldots,\sigma^n.$ Let us give one simple example. Recall that the measure $G$ is concentrated on the sphere $\|h\|=c$ and, for $q=c^2-\eps$, let $f_1(x) = t I(x \geq q)$ and $f_2 = \ldots = f_n =0.$ Then
$$
F(\sigma,\sigma^1,\ldots,\sigma^n) = t I(\sigma \cdot \sigma^1 \geq q)
$$
is a scaled indicator of a small neighborhood of $\sigma^1$ on the sphere $\|h\|=c$. If we denote by $W_1 = G(\sigma : \sigma \cdot \sigma^1 \geq q)$ the Gibbs weight of this neighborhood then the average in the denominator in (\ref{ch45main}) is equal to
$$
\la\exp F(\sigma,\sigma^1,\ldots,\sigma^n)\ra_{\hspace{-0.3mm}\mathunderscore}
=
W_1 e^t + 1-W_1.
$$
Suppose now that the function $\varPhi = I_A$ is an indicator of the event 
$$
A = \bigl\{(\sigma^1,\ldots, \sigma^n) : \sigma^1 \cdot \sigma^l <q \mbox{ for } 2\leq l\leq n \bigr\}
$$
that the replicas $\sigma^2,\ldots,\sigma^n$ are outside of this neighborhood of $\sigma^1$. Then, it is easy to see that
$$
\sum_{l=1}^{n} F_l(\sigma^l,\sigma^1,\ldots,\sigma^n)
=
t \e \la I(R_{1,2} \geq q)\ra
= : t\gamma
$$
and (\ref{ch45main}) becomes
\begin{equation}
\e \bigl\la I_A \bigr \ra =
\e\Bigl\la
I_A \smsp
\frac{e^{t\gamma}}
{(W_1 e^t + 1-W_1)^n}
\Bigr\ra,
\label{ch45mainEx}
\end{equation}
which may be viewed as a condition on the weight $W_1$ and the event $A$. This is just one artificial example, but the idea can be pushed much further and with some work one can obtain some very useful consequences about the structure of the measure $G$. One of these consequences is the following ``duplication property". 

\begin{figure}[t]
\centering
\psfrag{Sigma1}{$\sigma^1$}\psfrag{Sigma2}{$\sigma^2$}\psfrag{Sigman2}{$\sigma^{n-2}$}\psfrag{Sigman1}{$\sigma^{n-1}$}\psfrag{Sigman}{$\sigma^n$}
\includegraphics[width=0.55\textwidth]{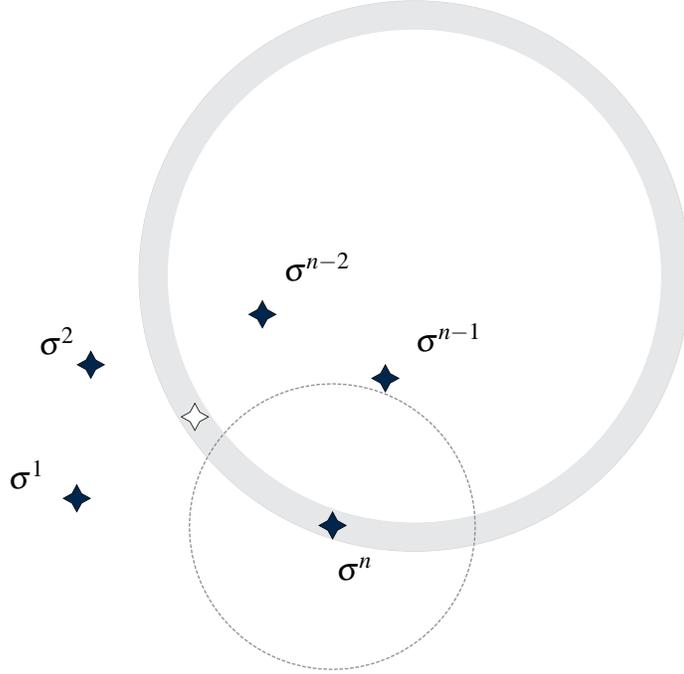}
\caption{\label{Fig2} {\it Duplication property.} The grey area corresponds to all the points on the sphere $\|h\|=c$ which are approximately at the same distance from the first $n-1$ replicas $\sigma^1,\ldots,\sigma^{n-1}$ as the last replica $\sigma^n$. Then the white point is a duplicate $\sigma^{n+1}$ of $\sigma^n$. It is in the grey area, so it is approximately at the same distances from the replicas $\sigma^1,\ldots, \sigma^{n-1}$ as $\sigma^n$, and it is at least as far from $\sigma^n$ as the closest of the first $n-1$ replicas, in this case $\sigma^{n-1}$.}
\end{figure}

Suppose that with positive probability over the choice of the measure $G$ we can sample $n$ replicas $\sigma^1,\ldots, \sigma^n$ from $G$ that are approximately at certain fixed distances from each other. Of course, since all replicas live on the same sphere, this can be expressed in terms of the overlaps, $R_{l,l'}\approx a_{l,l'}$, for some $n\times n$ matrix of constraints $A=(a_{l,l'})$. Let
$$
a_n^* = \max(a_{1,n},\ldots,a_{n-1,n})
$$
be the constraint corresponding to the closest point among $\sigma^1,\ldots,\sigma^{n-1}$ to the last replica $\sigma^n$ and suppose that the distance between them is strictly positive, $a_n^*<c^2$. Then, using the invariance property (\ref{ch45main}), one can show that with positive probability over the choice of $G$ one can sample $n+1$ replicas $\sigma^1,\ldots, \sigma^{n+1}$ from $G$ such that the distances between the first $n$ replicas are as above, $R_{l,l'}\approx a_{l,l'}$, and the new replica $\sigma^{n+1}$ duplicates $\sigma^n$ in the following sense (see Fig. \ref{Fig2}). First of all, it is approximately at the same distances from the replicas $\sigma^1,\ldots, \sigma^{n-1}$ as $\sigma^n$,
$$
R_{1,n+1} \approx a_{1,n},\ldots, R_{n-1,n+1} \approx a_{n-1,n},
$$
and, moreover, it is at least as far from $\sigma^n$ as the other closest replica, $R_{n,n+1} \lesssim a_n^*$. The motivation for this property becomes clear if we recall how the support of the Ruelle probability cascades was constructed in (\ref{ch43hs}). In that case it is obvious that one can always duplicate any replica with probability one, and not just in a weak sense described here. On the other hand, even this weak duplication property implies that the support of the measure $G$ is ultrametric with probability one. 

To see this, suppose that ultrametricity is violated and with positive probability three replicas can take values 
\begin{equation}
R_{1,2}\approx x, R_{1,3} \approx y \,\mbox{ and }\, R_{2,3}\approx z
\label{constraints}
\end{equation}
for some constraints $x<y\leq z <c^2$. 
Let us duplicate each replica in the above sense $m-1$ times, so that at the end we will have $n=3m$ replicas. Suppose that $$\{1,\ldots,n\} = I_1\cup I_2\cup I_3$$ is the partition such that $j\in I_j$ for $j\leq 3$, $|I_j| = m$ and each $I_j\setminus \{j\}$ is precisely the index set of duplicates of $\sigma^j$. Then, it should be almost obvious that
\begin{description}
\item[(a)] 
$R_{l,l'}\lesssim z$ for all $l\not = l'\leq n$,

\item[(b)] 
$R_{l,l'} \approx x$ if $l\in I_1, l'\in I_2$, $R_{l,l'} \approx y$ if $l\in I_1, l'\in I_3$ and $R_{l,l'} \approx z$ if $l\in I_2, l'\in I_3$.
\end{description}
Property (a) holds, because a new replica never get too close to the old replicas and the overlap will never exceed $z$, and property (b) holds, because, every time we duplicate a point, the distances to all other points will be the same, so the overlaps  between point in different groups $I_1, I_2, I_3$ will always be the same as the original constraints in (\ref{constraints}). This means that with positive probability we can find replicas $\sigma^1,\ldots, \sigma^n$ on the sphere of radius $c$ such that their overlaps satisfy properties (a) and (b). Let $\bar{\sigma}^j$ be the barycenter of  the set $\{\sigma^l : l\in I_j\}$. The condition (a) implies that  
$$
\|\bar{\sigma}^j\|^2 = \frac{1}{m^2}\sum_{l\in I_j} \|\sigma^l\|^2 + \frac{1}{m^2}\sum_{l\not = l'\in I_j} R_{l,l'}
\lesssim \frac{mc^2 + m(m-1) z}{m^2},
$$
and the condition (b) implies that $\bar{\sigma}^1\cdot \bar{\sigma}^2 \approx x$, $\bar{\sigma}^1\cdot \bar{\sigma}^3 \approx y$ and $\bar{\sigma}^2\cdot \bar{\sigma}^3 \approx z$. Hence,
$$
\|\bar{\sigma}^2-\bar{\sigma}^3\|^2 = \|\bar{\sigma}^2\|^2 + \|\bar{\sigma}^3\|^2 
- 2\bar{\sigma}^2\cdot \bar{\sigma}^3 \lesssim \frac{2(c^2 -z)}{m}
$$
and $0< b-a \approx \bar{\sigma}^1 \cdot \bar{\sigma}^3 -\bar{\sigma}^1\cdot \bar{\sigma}^2 \lesssim K m^{-1/2}.$ We arrive at contradiction by letting $m\to\infty$. The conclusion is that if the measure $G$ satisfies the Ghirlanda-Guerra identities then its support must be ultrametric with probability one. Therefore, as was explained above, under a small perturbation of the Hamiltonian that yields the Ghirlanda-Guerra identities, all possible limits of the Gibbs measure can be identified with the Ruelle probability cascades. Moreover, we will see below that for ``generic" mixed $p$-spin models no perturbation is necessary and the limit of the Gibbs measure is unique.

\section{Consequences of the Parisi formula}

Once we know the Parisi formula for the free energy in the mixed $p$-spin models, some results can be extended and strengthened. 

\paragraph{Universality in the disorder.}
First of all, one can prove the universality in the disorder and show that the Parisi formula (\ref{ch30Parisi}) still holds if the Gaussian random variables $(g_{ij})$ in the Hamiltonian (\ref{SKH}) are replaced by i.i.d. random variables $(x_{ij})$ from any other distribution, as long as
\begin{equation}
\e x_{11} = 0, \e x_{11}^2 =1 \,\mbox{ and }\, \e |x_{11}|^3 <\infty.
\label{ch38disorder}
\end{equation}
This was proved by Carmona and Hu in \cite{CarmonaHu}, who generalized an earlier result of Talagrand \cite{TalGB} in the case of the Bernoulli disorder. The proof is based on the following interpolation between the two Hamiltonians for $0\leq t\leq 1$,
\begin{equation}
H_{N,t}(\sigma) = \frac{1}{\sqrt{N}} \sum_{i,j =1}^N \bigl(\sqrt{t}x_{ij} + \sqrt{1-t} g_{ij}\bigr)\sigma_i \sigma_j,
\label{ch38SKHInt}
\end{equation}
and the estimates of the derivative of the free energy along this interpolation using some approximate integration by parts formulas. The same result was also proved in \cite{CarmonaHu} for the $p$-spin model.

\paragraph{Generic mixed $p$-spin models.}
In another direction, we can say more about the thermodynamic limit in the case of the so-called generic mixed $p$-spin models whose Hamiltonian (\ref{mixedH}) contains sufficiently many pure $p$-spin terms (\ref{mixedp}), so that the following condition is satisfied:
\begin{enumerate}
\item[(G)] linear span of constants and power functions $x^p$ corresponding to $\beta_p \not = 0$ is dense in\\ $(C[-1,1],\|\cdot\|_\infty)$. 
\end{enumerate}
The reason for this is that each pure $p$-spin term contains some information about the $p$th moment of the overlap and the condition (G) allows us to confirm the Parisi ansatz for the Gibbs measure in the thermodynamic limit without the help of the perturbation term in (\ref{ch31mixedHpert}), so the result becomes more pure, in some sense. Moreover, in this case we can show that the asymptotic Gibbs measure is unique. This can be seen in two steps. Let us denote by $\PP = \PP\bigl((\beta_p)\bigr)$ the infimum on the right-hand side of (\ref{ch30Parisi}) and let $\M$ be the set of all limits over subsequences of the distribution of the overlap $R_{1,2}$ of two spin configurations sampled from the Gibbs measure $G_N$ corresponding to the Hamiltonian (\ref{mixedH}). It was proved by Talagrand in \cite{PM} that, for each $p\geq 1$, the Parisi formula is differentiable with respect to  $\beta_p$ and 
\begin{equation}
\frac{\partial \PP}{\partial \beta_p} = 
\beta_p\Bigl(1-\int\! q^p \smsp d\zeta(q) \Bigr)
\label{ch36derP}
\end{equation}
for all $\zeta\in \M$. If $\beta_p \not = 0$, then this implies that all the limits $\zeta\in \M$ have the same $p$th moment and the condition (G) then implies that $\M = \{\zeta_0\}$ for some unique distribution $\zeta_0$ on $[-1,1]$. 
As a second step, one can prove the convergence of the entire overlap array $(R_{l,l'})_{l,l'\geq 1}$ in distribution as follows. As a consequence of the differentiability of the Parisi formula, it was proved in \cite{PGGmixed} that, whenever $\beta_p\not = 0$, the Ghirlanda-Guerra identities
\begin{equation}
\e  \bigl\la f R_{1,n+1}^p \bigr\ra 
=
 \frac{1}{n}\e \bigl\la f \bigr\ra \smsp \e\bigl\la R_{1,2}^p\bigr\ra 
 +
 \frac{1}{n}\sum_{l=2}^{n}\e \bigl\la f R_{1,l}^p\bigr\ra
\label{ch31GGlimitAgain}
\end{equation}
for the $p$th moment of the overlap hold in the thermodynamic limit in a strong sense, for the Gibbs measure $G_N$ corresponding to the original Hamiltonian (\ref{mixedH}) without the perturbation term (\ref{ch31mixedHpert}). The condition (G) then, obviously, implies that the general Ghirlanda-Guerra identities (\ref{ch31GGlimit2}) also hold. As we explained above, the Ghirlanda-Guerra identities imply ultrametricity and, as a result, the distribution of the entire overlap array can be uniquely determined by the distribution of one overlap. Since this distribution $\zeta_0$ is unique, the distribution of the entire overlap array under $\e G_N^{\otimes\infty}$ also has a unique limit, so the asymptotic Gibbs measure is unique.  Notice also that, by Talagrand's positivity principle, the distribution $\zeta_0$ is, actually, supported on $[0,1]$.
Finally, in this case one can show using the Aizenman-Sims-Starr scheme (\ref{ch12AS2repr}) that the limit of the free energy is equal to $\PP(\zeta_0)$, where we understand that the definition of the Parisi functional $\PP(\zeta)$ is extended to all distributions on $[0,1]$ by continuity. Thus, the infimum $\inf_{\zeta} \PP(\zeta)$ in the Parisi formula (\ref{ch30Parisi}) is achieved on the asymptotic distribution of the overlap, $\zeta_0$. The functional $\PP(\zeta)$ is conjectured to be convex in $\zeta$ (see \cite{PConvex} for a partial result)  and, if true, this would imply that $\zeta_0$ is the unique minimizer of $\PP(\zeta)$. Convexity of $\PP(\zeta)$ would also give a more direct approach to describing the high temperature region, which was done by Talagrand in \cite{SG} (see also \cite{SG2}).



\begin{thebibliography}{999}


\bibitem{AC} Aizenman, M., Contucci, P.: On the stability of the quenched state in mean-field spin-glass models. J. Statist. Phys. \textbf{92}, no. 5-6, 765--783 (1998)

\bibitem{AS2} Aizenman, M., Sims, R., Starr, S.L.: An extended variational principle for the SK spin-glass model. Phys. Rev. B. \textbf{68}, 214403 (2003)

\bibitem{AA} Arguin, L.-P., Aizenman, M.: On the structure of quasi-stationary competing particles systems. Ann. Probab. \textbf{37}, no. 3, 1080--1113 (2009)

\bibitem{ACh} Arguin, L.-P., Chatterjee, S.: Random overlap structures: properties and applications to spin glasses. Probab. Theory Related Fields (2012) doi: 10.1007/s00440-012-0431-6

\bibitem{Bolthausen} Bolthausen, E., Sznitman, A.-S.: On Ruelle's probability cascades and an abstract cavity method. Comm. Math. Phys. \textbf{197}, no. 2, 247--276 (1998) 

\bibitem{Bovier}  Bovier, A., Kurkova, I.: Derrida's generalized random energy models. I. Models with finitely many hierarchies.  Ann. Inst. H. Poincar\'e Probab. Statist. \textbf{40},  no. 4, 439--480 (2004)  

\bibitem{CarmonaHu} Carmona, P., Hu, Y.: Universality in Sherrington-Kirkpatrick's spin glass model. Ann. Inst. H. Poincar\'e Probab. Statist. \textbf{42}, no. 2, 215--222  (2006)

\bibitem{CGSS} Contucci, P., Giardina, C.: Spin-glass stochastic stability: a rigorous proof.  Ann. Henri Poincar\'e \textbf{6}, no. 5, 915--923 (2005) 

\bibitem{deDH} de Dominicis, C., Hilhorst, H.: Random (free) energies in spin glasses, J. Phys. Lett. \textbf{46}, L909-L914 (1985)

\bibitem{DerridaREM1} Derrida, B.: Random-energy model: limit of a family of disordered models. Phys. Rev. Lett. \textbf{45}, no. 2, 79--82 (1980)

\bibitem{DerridaREM2} Derrida, B.: Random-energy model: an exactly solvable model of disordered systems. Phys. Rev. B (3) \textbf{24}, no. 5, 2613--2626 (1981)

\bibitem{DerridaGREM} Derrida, B.: A generalization of the random energy model that includes correlations between the energies. J. Phys. Lett. \textbf{46}, 401--407 (1985) 

\bibitem{DerridaGREM2} Derrida, B., Gardner, E.: Solution of the generalised random energy model. J. Phys. C \textbf{19}, 2253--2274 (1986)

\bibitem{DerridaToulouse} Derrida, B., Toulouse, G.: Sample to sample fluctuations in the random energy model. J. Phys. Lett. \textbf{46}, L223--L228 (1985)

\bibitem{DS} Dovbysh, L. N., Sudakov, V. N.:  Gram-de Finetti matrices. Zap. Nauchn. Sem. Leningrad. Otdel. Mat. Inst. Steklov.  \textbf{119}, 77--86  (1982)

\bibitem{GG} Ghirlanda, S., Guerra, F.: General properties of overlap probability distributions in disordered spin systems. Towards Parisi ultrametricity.  J. Phys. A  \textbf{31}, no. 46, 9149--9155 (1998) 

\bibitem{GuerraToninelli} Guerra, F., Toninelli, F.L.: The thermodynamic limit in mean field spin glass models. Comm. Math. Phys. \textbf{230}, no. 1, 71--79 (2002)

\bibitem{Guerra} Guerra, F.: Broken replica symmetry bounds in the mean field spin glass model. Comm. Math. Phys. {\bf 233}, no. 1, 1--12 (2003)

\bibitem{king} Kingman, J. F. C.:  Poisson Processes. Oxford University Press, New York (1993)

\bibitem{M1} M\'ezard, M., Parisi, G., Sourlas, N., Toulouse, G., Virasoro, M.A.: On the nature of the spin-glass phase. Phys. Rev. Lett. \textbf{52}, 1156 (1984) 

\bibitem{M2} M\'ezard, M., Parisi, G., Sourlas, N., Toulouse, G., Virasoro, M.A.: Replica symmetry breaking and the nature of the spin-glass phase. J. de Physique \textbf{45}, 843 (1984)

\bibitem{MPV} M\'ezard, M., Parisi, G., Virasoro, M.A.: Spin Glass Theory and Beyond. World Scientific Lecture Notes in Physics, 9. World Scientific Publishing Co., Inc., Teaneck, NJ (1987) 

\bibitem{PConvex} Panchenko, D.: A question about the Parisi functional. Elect. Comm. in Probab. \textbf{10}, 155--166 (2005)

\bibitem{PDS} Panchenko, D.: On the Dovbysh-Sudakov representation result. Electron. Comm. in Probab.  \textbf{15}, 330--338 (2010)

\bibitem{PGG} Panchenko, D.: A connection between Ghirlanda-Guerra identities and ultrametricity.  Ann. of Probab. \textbf{38}, no. 1, 327--347 (2010)

\bibitem{PGGmixed} Panchenko, D.: The Ghirlanda-Guerra identities for mixed $p$-spin model. C.R. Acad. Sci. Paris, Ser. I \textbf{348}, 189--192 (2010) 

\bibitem{PGG2} Panchenko, D.: Ghirlanda-Guerra identities and ultrametricity: An elementary proof in the discrete case. C. R. Acad. Sci. Paris, Ser. I \textbf{349}, 813--816 (2011)

\bibitem{ACGG} Panchenko, D.: A unified stability property in spin glasses.  Comm. Math. Phys. \textbf{313}, no. 3, 781--790 (2012) 

\bibitem{PUltra} Panchenko, D.: The Parisi ultrametricity conjecture. Ann. of Math. (2). \textbf{177}, no. 1, 383--393 (2013)

\bibitem{PPF} Panchenko, D.: The Parisi formula for mixed p-spin models. To appear in Ann. of Probab. (2011)

\bibitem{SKmodel} Panchenko, D.: The Sherrington-Kirkpatrick Model. Springer Monographs in Mathematics. Springer-Verlag, New York, xii+158 pp. ISBN:  978-1-4614-6288-0 (2013)

\bibitem{Parisi79} Parisi, G.: Infinite number of order parameters for spin-glasses. Phys. Rev. Lett. \textbf{43}, 1754--1756 (1979)

\bibitem{Parisi} Parisi, G.: A sequence of approximate solutions to the S-K model for spin glasses. J. Phys. A \textbf{13}, L-115 (1980) 

\bibitem{Parisi83} Parisi, G.: Order parameter for spin glasses. Phys. Rev. Lett. \textbf{50}, 1946 (1983) 

\bibitem{Pastur} Pastur, L. A., Shcherbina, M. V.: Absence of self-averaging of the order parameter in the Sherrington-Kirkpatrick model. J. Statist. Phys. \textbf{62}, no. 1--2, 1--19 (1991)

\bibitem{Ruelle} Ruelle, D.: A mathematical reformulation of Derrida's REM and GREM.  Comm. Math. Phys.  {\bf 108},  no. 2, 225--239 (1987)

\bibitem{SK} Sherrington, D., Kirkpatrick, S.: Solvable model of a spin glass. Phys. Rev. Lett. {\bf 35}, 1792--1796 (1975)

\bibitem{TalGB} Talagrand, M.: Gaussian averages, Bernoulli averages, and Gibbs' measures. Random Struct. Algorithms \textbf{21} (3-4), 197--204 (2002)

\bibitem{SG} Talagrand, M.:  Spin Glasses: a Challenge for Mathematicians.  Ergebnisse der Mathematik und ihrer Grenzgebiete. 3. Folge A Series of Modern Surveys in Mathematics, Vol. 43. Springer-Verlag (2003) 

\bibitem{T2} Talagrand, M.:  On Guerra's broken replica-symmetry bound.  C. R. Math. Acad. Sci. Paris {\bf 337}, no. 7, 477--480 (2003)

\bibitem{PM} Talagrand, M.: Parisi measures.  J. Funct. Anal. \textbf{231}, no. 2, 269--286 (2006)

\bibitem{TPF} Talagrand, M.: The Parisi formula. Ann. of Math. (2) \textbf{163}, no. 1, 221--263 (2006)

\bibitem{Tal-New} Talagrand, M.: Construction of pure states in mean-field models for spin glasses. Probab. Theory Relat. Fields \textbf{148}, no. 3-4, 601--643 (2010)

\bibitem{SG2} Talagrand, M.: Mean-Field Models for Spin Glasses. Ergebnisse der Mathematik und ihrer Grenzgebiete. 3. Folge A Series of Modern Surveys in Mathematics, Vol. 54, 55. Springer-Verlag (2011)






\end{thebibliography}
\end{document}